\documentclass[10pt,letterpaper]{article}
\usepackage{opex3}

\begin{document}

\title{Experimental generation of 6 dB continuous variable entanglement from
a nondegenerate optical parametric amplifier}
\author{Yu Wang, Heng Shen, Xiaoli Jin, Xiaolong Su, Changde Xie, and Kunchi
Peng}
\address{State Key Laboratory of Quantum Optics and Quantum
Optics Devices, Institute of Opto-Electronics, Shanxi University,
Taiyuan, 030006, People's Republic of China}
\email{suxl@sxu.edu.cn}

\begin{abstract}
We experimentally demonstrated that the quantum correlations of amplitude
and phase quadratures between signal and idler beams produced from a
non-degenerate optical parametric amplifier (NOPA) can be significantly
improved by using a mode cleaner in the pump field and reducing the phase
fluctuations in phase locking systems. Based on the two technical
improvements the quantum entanglement measured with a two-mode homodyne
detector is enhanced from $\sim $ 4 dB to $\sim $\ 6 dB below the quantum
noise limit using the same NOPA and nonlinear crystal.
\end{abstract}






\section{Introduction}

Continuous variable (CV) quantum entanglement between amplitude and phase
quadratures of optical fields has been intensively investigated. The optical
CV entangled states have been utilized in quantum information science to
realize deterministic quantum teleportation of coherent states \cite%
{Fur,Yon2004,Bowen2003}, unconditional dense coding quantum communication
\cite{Li,Jing}, entanglement swapping \cite{Jia}, quantum key distribution
\cite{Ben,Su2009}, and so on. Recently, it has been proved that CV
entanglement can be applied to develop quantum computation and universal
quantum information process \cite%
{Menicucci2006,Tan2008,Yoshikawa2008,Fur2007}.

At the beginning of 1990s, Kimble's group experimentally generated a pair of
CV entangled optical beams by type-II down-conversion in a subthreshold
nondegenerate optical parametric amplifier (NOPA) and demonstrated the
Einstein-Podolsky-Rosen (EPR) paradox firstly in CV regime \cite{Pereira,
Ou1992}. The quantum correlations of amplitude and phase quadratures between
output signal and idler beams observed in their experiment were $3.6\pm 0.2$
dB below the quantum noise limit (QNL) normalized to the vacuum noise of the
corresponding optical field. Almost ten years later, the bright EPR
entangled beams with $\sim $ 4 dB amplitude (phase) correlation and phase
(amplitude) anticorrelation were generated from a NOPA with injected signals
operating at amplification (deamplification) \cite{Li,
Zhang2000,Bowen20032,Laurat2005}. Intense EPR entangled beams were produced
from optical parametric oscillators (OPOs) operating above threshold by
several groups recently \cite{Villar2005,Su2006,Jing2006}. However, the
quadrature-phase correlations of these intense beams were worse than that
obtained from NOPAs below threshold. Another efficiently and extensively
used scheme generating CV entangled states is to interfere two single-mode
squeezed beams on a $1:1$ beamsplitter \cite{Fur,Yon2004}. The two squeezed
beams should have identical frequency to achieve stable interference. In
these experiments, they were produced from two degenerate optical parametric
amplifiers (DOPAs) with identical type-I nonlinear crystal pumped by a laser
to ensure their optical frequency being the same. Recent years, the
single-mode squeezed light beams of $\sim $ 9 dB and $\sim $ 10 dB were
generated from a DOPA with a PPKTP (periodically poled KTiOPO4) crystal and
a monolithic DOPA made by MgO:LiNbO3 crystal, respectively, based on some
technical improvements \cite{Tak,Vah}. It is exciting that CV four-mode
cluster states with the multi-mode quadrature correlations of $\sim $ 6 dB
below the QNL have been experimentally observed recently \cite{Yukawa2008}.
In the system of Ref. \cite{Yukawa2008}, the four initial single-mode
squeezed states were generated from four DOPOs pumped by a Ti:sapphire
laser. However, the correlations of EPR beams from NOPAs still keep at $\sim
$ 4 dB so far. Under the motivation of the great achievement on DOPAs, we
followed the ideas in Ref. \cite{Tak, Vah} to implement some technical
improvements on the system for EPR entangled state generation from a NOPA.
At first, the amplitude and phase noises of the pump laser of the NOPA were
reduced to the QNL level by using a specifically designed mode cleaner (MC).
Second, the phase fluctuations in all phase locking systems were controlled
detailedly. Based on above improvements the quantum correlations of
quadrature components between EPR entangled beams generated from a NOPA were
increased from $\sim $ 4 dB to $\sim $ 6 dB below the QNL. The experiment
results show that there is potential to further enhance the CV entanglement
of output optical beams from NOPAs.

\section{Experimental setup}

The schematic of the EPR entangled state generation system is shown in
Figure 1. The laser is a continuous wave intra-cavity frequency-doubled and
frequency stabilized Nd:YAP/KTP (Nd-doped YAlO3 perovskite / potassium
titanyl phosphate) laser (Yuguang Co. Ltd., F-VIB). The harmonic wave of 2 W
at 540 nm and the fundamental wave of 0.8 W at 1080 nm are generated
simultaneously from the laser. The output green and infrared lasers are
separated by a mirror coating with high reflectivity for 540 nm and high
transmission for 1080 nm, and then are used as the pump field and the
injected signal of the NOPA, respectively. The traveling-wave resonator
placed in the laser beam of 1080 nm serves as the optical low-pass filter of
noises and the spatial mode cleaner (MC1). The finesse and the linewidth of
MC1 for 1080 nm are 700 and 1 MHz, respectively. An important improvement
with respect to our previously experimental system \cite{Li,Zhang2000} is
that a mode cleaner (MC2) with a finesse of 1000 and linewidth of 600 kHz
for 540 nm is added in the pump laser before the NOPA. MC2 is also a
traveling-wave resonator consisting of three mirrors with the total cavity
length of 520 mm. Both MC1 and MC2 are temperature-controlled to improve the
mechanic stability. The MC2 is adjusted to resonating with the second
harmonic laser field via a Pound-Drever-Hall locking scheme with a phase
modulation frequency at 5.8 MHz \cite{Pound}. We found that MC2 not only can
significantly improve the quality of spatial distribution of the pump laser
but also can reduce its phase fluctuation.

\begin{figure}[tbp]
\begin{center}
\includegraphics[width=10cm,height=4.5cm]{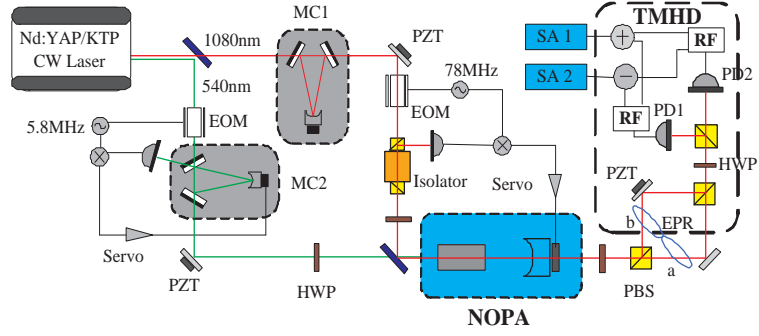}
\end{center}
\caption{(Color online) Schematic of the experimental setup. MC1-2: mode
cleaner, PZT: pieoelectric transducer, Servo: servo amplifier circuit for
feedback system, EOM: electro-optical modulator, HWP: half wave-plate, NOPA:
nondegenerate optical parametric amplifier, PBS: polarization beamsplitter,
TMHD: two-mode homodyne detector for continuous variable entanglement,
PD1-2: photodiode detector (ETX500 InGaAs), RF: radio frequency power
splitter, $\oplus $: positive power combiner; $\ominus $: negative power
combiner, SAs: spectrum analyzers.}
\end{figure}

The NOPA consists of an $\alpha $-cut type-II KTP ($3\times 3\times 10$ mm$%
^{3}$) crystal and a concave mirror. The front face of KTP is coated to be
used as the input coupler (the transmission 99.8\% at 540 nm wavelength and
0.04\% at 1080 nm) and the other face of which is coated with the dual-band
antireflection at both 540 nm and 1080 nm. A concave mirror of 50 mm
curvature radius, which is coated with transmission $T=5.2\%$ at 1080 nm and
high reflectivity at 540 nm, is used as the output coupler of the EPR beams
at 1080 nm. The measured finesse at 1080 nm is 117, so the intra-cavity loss
is $L=0.17\%$. The cavity length of the NOPA is 54 mm and the waist size
radius of the infrared field at the input face of KTP crystal is
approximately 38 $\mu $m. The concave mirror is mounted on a piezoelectric
transducer (PZT) for locking actively the cavity length of NOPA on resonance
with the injected signal at 1080 nm via a Pound-Drever-Hall locking scheme
with the phase modulation frequency of 78 MHz \cite{Pound}. The temperature
of KTP crystal are carefully controlled at $63%
{{}^\circ}%
$C to satisfy the type-II phase match condition. Through a parametric down
conversion process of type-II phase match, a bright EPR beam with
anticorrelated amplitude quadratures and correlated phase quadratures can be
produced from a NOPA operating in the state of deamplification, where the
relative phase between pump field and the injected signal is locked to $\pi $
\cite{Li}. The correlation variances of amplitude ($\hat{X}$) and phase ($%
\hat{Y}$) quadratures of the two EPR beams $\hat{a}$\ and $\hat{b}$ are
given by $\langle \delta ^{2}(\hat{X}_{a}+\hat{X}_{b})\rangle =\langle
\delta ^{2}(\hat{Y}_{a}-\hat{Y}_{b})\rangle =2e^{-2r}$, where $r$ is the
correlation parameter which depends on the strength and the time of
parametric interaction in NOPA \cite{Su2009}. The values of $r=0$ and $%
r\rightarrow +\infty $ correspond to no correlation and ideal correlation,
respectively.\ The EPR beams, which have identical frequency with the
injected signal and the orthogonal polarization with each other, are
separated by a polarization beam splitter (PBS). The quantum correlations
between amplitude and phase quadratures of the two EPR beams are measured by
a two-mode homodyne detector (TMHD) for continuous variable entanglement
\cite{ZhangJ}, which consists of two PBS, a half wave-plate (HWP), two
ETX-500 photodiode detectors (PD1 and PD2) and two radio frequency power
splitters (RFs). The detector requires to combine the two entangled optical
modes on a beamsplitter, thus it is not able to be used for verifying the
entanglement of two spatially separated modes and also can not be applied in
the measurement of the squeezed vacuum states due to that it can only work
with bright signal and idler beams.\textbf{\ }Using a PZT placed in one of
the detected two beams, we lock the relative phase between two EPR beams to $%
\pi /2$. The output photo-currents from the radio frequency power splitter
are combined by the positive power combiner ($\oplus $) and the negative
power combiner ($\ominus $) to obtain the noise powers of amplitude sum and
phase difference, respectively. Then they are recorded by spectrum analyzers
1 and 2, respectively. To determine the corresponding QNL, we blocked the
pump field of NOPA and locked the NOPA to resonate with the injected seed
light. Adjusting the power of the injected seed light to make the intensity
of the output coherent state at 1080 nm exactly equals that of the generated
entangled state, in this case the measured variances are the corresponding
QNL.

\section{Experimental results and discussion}

\begin{figure}[tbp]
\begin{center}
\includegraphics[width=10cm,height=10cm]{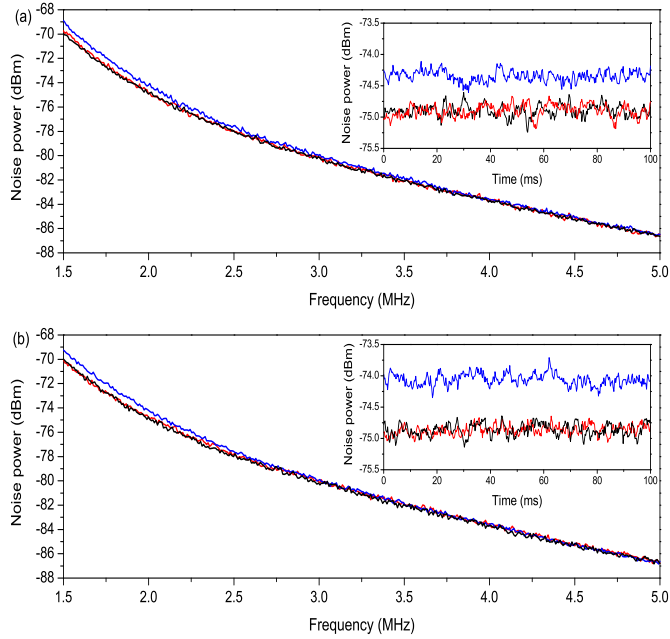}
\end{center}
\caption{(Color online) The measured noise powers of amplitude (a) and phase
(b) quadratures of pump laser. Black lines: QNL, red lines: noise of pump
laser with MC2, blue lines: noise of pump laser without MC2. Insets: the
noises of amplitude (a) and phase (b) quadratures at 2 MHz. The parameter of
spectrum analyzer: RBW: 30 kHz, VBW: 100 Hz.}
\end{figure}

To analyze the effect of the pump noises we measured the amplitude
and phase noise power spectra of the pump laser at 540 nm with and
without the use of MC2, which are shown in Fig. 2 (a) and (b)
respectively. The amplitude (a) and the phase (b) noises in Fig. 2
were measured by another balanced homodyne system, which is not
shown in Fig. 1. The used laser powers for the local oscillation
beam and the signal beam for the homodyne measurement are 5.7 mW and
30 $\mu $W, respectively. The measurements for both cases with and
without the use of MC2 were implemented before NOPA. The photodiode
used in the balanced homodyne system is S5973-02 (Hamamatsu). In the
measurements, the noises of the used local beams have been filtered
to the QNL and the resolution bandwidth (RBW) and video bandwidth
(VBW) are 30 kHz and 100 Hz, respectively.\textbf{\ }In Fig. 2, the
black, red and blue lines correspond to the QNL, the noise power of
the output coherent state with and without the use of MC2,
respectively. We can see that if MC2 is not used both amplitude and
phase noises are higher than the QNL from 1.5 MHz to 4 MHz. These
excess noise in the pump laser come from the relaxation oscillation
of laser and they decrease along with the increase of the analysis
frequency gradually to the QNL after 4 MHz. The noise levels of
amplitude and phase quadratures of the pump laser are 0.5 dB and 1
dB above the QNL at 2 MHz, respectively [see the insets in Fig. 2
(a) and (b)]. After the use of MC2, the amplitude and phase noises
are reduced to the QNL at 2 MHz, where the correlation variances
will be measured.

It has been proved that the relative phase fluctuations in homodyne
measurement significantly influence the measured correlations \cite{Tak}. To
reduce the relative phase fluctuations the stack PZT (Piezomechanik Gmbh,
HPSt) is chosen and the electronic feedback system is optimized. At first,
the modulation frequency and the modulation depth are adjusted to give an
error signal with higher signal-to-noise ratio. Then a low noise highvoltage
amplifier (Yuguang Co. Ltd. YG2009A-350V) with a lower output voltage noise
of 6 mV$_{pp}$ is used for amplifying the error signal. After these
improvements, the phase fluctuation of the phase locking system reach 1.8$%
{{}^\circ}%
$.

To produce the entangled states of light with anticorrelated amplitude
quadratures and correlated phase quadratures, the NOPA is pumped with a
green laser of 170 mW which is 60 mW below the oscillation threshold of 230
mW and is operated at deamplification \cite{Li,Zhang2000}. In this case, the
intensity of the output EPR entangled beams is $\sim $ 80 $\mu $W and the
parametric gain is $g=25$. The escape efficiency and the cavity linewidth of
the NOPA are $\eta _{esc}=T/(T+L)=96.8\%$ and $\Delta \nu =23.7$ MHz,
respectively. The detection efficiency and the mode-matching efficiency of
the detection system are 90\% and 99.9\%, respectively.

\begin{figure}[tbp]
\begin{center}
\includegraphics[width=12cm,height=6cm]{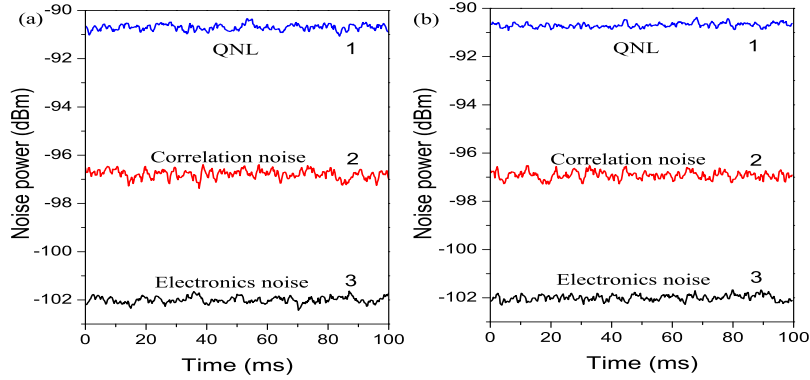}
\end{center}
\caption{(Color online) The measured noise powers of amplitude sum (a) and
phase difference (b) at 2 MHz. The parameter of spectrum analyzer: RBW: 30
kHz, VBW: 100 Hz.}
\end{figure}

The measured noise powers of the amplitude sum (a) and the phase difference
(b)\ of the generated EPR beams are shown in Fig. 3. Pairs of QNLs,
correlated noise and electronic noise are measured simultaneously by two
spectrum analyzers with the following parameters: measurement frequency of 2
MHz in zero span mode, RBW of 30 kHz, and VBW of 100 Hz. Trace (1) is the
QNL, which corresponds to the noise level of coherent state with a power of
80 $\mu $W. Traces (2) in Fig. 3 (a) and (b) are the measured correlation
variances of the amplitude sum $\langle \delta ^{2}(\hat{X}_{a}+\hat{X}%
_{b})\rangle $ and the phase difference $\langle \delta ^{2}(\hat{Y}_{a}-%
\hat{Y}_{b})\rangle $ respectively, which are $6.08\pm 0.18$ dB and $6.22\pm
0.16$ dB below the corresponding QNL. The electronic noise of detector
[trace (3)] is approximately 11 dB below the QNL. Considering the influence
of electronics noise, the actual quantum correlation should be $\langle
\delta ^{2}(\hat{X}_{a}+\hat{X}_{b})\rangle =7.30\pm 0.18$ dB and $\langle
\delta ^{2}(\hat{Y}_{a}-\hat{Y}_{b})\rangle =$ $7.50\pm 0.16$ dB below the
corresponding QNL. The generated bright EPR entangled beams satisfy the
inseparable criteria \cite{Duan}:\textbf{\ }$\langle \delta ^{2}(\hat{X}_{a}+%
\hat{X}_{b})\rangle +\langle \delta ^{2}(\hat{Y}_{a}-\hat{Y}_{b})\rangle =$%
\textbf{\ }$0.485<2$, which corresponds to the observed entanglement of 6 dB
without subtracting electronics noise.

In most theoretical discussions the noises of the pump lasers of NOPA are
assumed at the QNL without excess noises. However, in practical cases the
excess noises usually exist in the pump lasers which will enter in the
output fields of NOPA during the parametric conversion and thus decrease its
entanglement. Of course, the factors limiting entanglement in NOPA devices
are of variety. Generaly, the entanglement from NOPA are mainly limited by
the quality of nonlinear crystal (nonlinear coefficient and losses), escape
efficiency and cavity linewidth\ of NOPA, detection efficiency,
mode-matching efficiency of detection system, and phase fluctuation of phase
locking system \cite{Tak}. Towards further increasing the entanglement for
our system, the currently limiting factors mainly are the lower escape
efficiency of the optical cavity of the NOPA and the lower detection
efficiency of the detection system. If we increase the escape efficiency
from present 96.8\% to 98.6\% ($T=12\%$) and the detection efficiency from
present 90\% to 99\%, over 10 dB entanglement will be expected. Besides,
observing high entanglement the electronic noise background of the detection
system has to be reduced to a large extent, for that the present TMHD should
be replaced by the normal balanced homodyne detector \cite{Ou1992}. At last,
we should mention that the infinite entanglement can not be obtained because
it requires infinite energy \cite{Coc}.

\section{Conclusion}

For the conclusion, we obtained CV optical entangled states with amplitude
and phase quadrature quantum correlations of $\sim $ 6 dB, which are the
highest correlations generated from NOPA devices so far to the best of our
knowledge. Comparing with our previous NOPA systems \cite{Li, Zhang2000},
two technical improvements are made: 1. Both amplitude and phase noises of
the pump laser of NOPA are reduced to the QNL level with a suitable mode
cleaner. 2. The relative phase fluctuations of phase locking systems are
minimized. Our experiment shows that the implementation of NOPAs on
entanglement generation can be significantly enhanced by means of some
technical improvements on the pump laser and the phase locking systems even
using the same NOPA configuration and nonlinear crystal.

\section*{Acknowledgments}

This research was supported by the NSFC (Grants No. 60736040 and 10804065),
NSFC Project for Excellent Research Team (Grant No. 60821004), National
Basic Research Program of China (Grant No. 2007BAQ03918).

\end{document}